\documentstyle[12pt]{article}

\newcommand{\bea}{\begin{eqnarray}}
\newcommand{\eea}{\end{eqnarray}}

\title{
\textbf{External leg amputation in conformal invariant three-point function}}
\author{
Indrajit Mitra\footnote{E-mail: indrajit.mitra@saha.ac.in, imphys@caluniv.ac.in}\\\\
Department of Theoretical Physics,\\
Indian Association for the Cultivation of Science,\\
2A \& 2B, Raja S. C. Mullick Road, Kolkata 700032, India\\
and\\
Department of Physics, University of Calcutta,\\
92 A.P.C. Road, Kolkata 700009, India\footnote{Present address}}

\date{}
\begin{document}
\maketitle
\begin{abstract}
Amputation of external legs is carried out explicitly for the conformal invariant three-point 
function involving two spinors and one vector field. Our results are 
consistent with the general result that amputing an external leg in a 
conformal invariant Green function replaces a field by its conformal partner in the Green function. 
A new star-triangle relation, involving two spinors and one vector field, 
is derived and used for the calculation.  
\end{abstract}

\section{Introduction}
\label{intro}
This work is concerned with the amputation of external legs in conformal invariant Green functions
in Euclidean space with general number of dimensions.
Various aspects of CFT (conformal field theory) in $D$ dimensions have been reviewed in 
Refs.\ \cite{fradkin1}, \cite{fradkin2} and \cite{book}. We consider conformal invariant Green 
functions involving spinors and vector fields, which are relevant for the
infrared limit of massless  QED$_3$ \cite{mrs1, mrs2}, and for conformal QED$_4$
\cite{palchik, comm}. Some other areas which use $D$-dimensional CFT's with fields of non-zero
spin are ${\cal N}=4$ supersymmetric Yang-Mills theory, conformal windows of gauge theories,
and unparticle physics \cite{georgi, georgi'}.  

A conformal invariant Green function includes the external legs, but amputed Green
functions are easier to calculate, because we do not have to integrate (in position space) over
the extra vertices. This provides a motivation for studying amputation.
Moreover, the conformal partial wave expansion \cite{fradkin1, fradkin2, book, parisi1, parisi2}
involves amputed Green function. This expansion expresses the contribution of the various 
quasi-primary
fields to the product of two field operators at arbitrary separation. From this, one can find
the contributions of the quasi-primary
fields to the four-point function. A recent work which uses the conformal partial wave expansion
and the amputed three-point function is Ref.\ \cite{georgi'}.

Formally, amputation of an external leg in a Green function in $D$-dimensional CFT replaces a field of
scale dimension $d$ by its conformal partner, having scale dimension $D-d$
\cite{fradkin1, fradkin2, book, parisi1, parisi2}. However, only an
explicit calculation can determine the coefficient which comes with the amputed Green function.
The simplest calculation is that for scalar field theory. Compared to this, the case of
massless Yukawa theory \cite{fradkin1}
is more involved, as there are more than one invariant structures
for a given Green function. But we find that amputation in the
case involving spinors and vector field is much more complicated. For amputing one of the two invariant
structures in this case, we have to derive and make use of a new star-triangle relation.

Indeed, the techniques of calculation developed in this paper can be useful in other areas which
involve evaluation of massless Feynman integrals, like ${\cal N}=4$ Yang-Mills theory \cite{cvetic}.
The star-triangle relation involving scalar fields \cite{deramo, sym}
(referred to as the D'EPP formula in
Ref.\ \cite{georgi'}) has wide-ranging applications: see Ref.\ \cite{isaev1} and references
therein. It is expected that the analogous relation, involving two spinors and
one vector field, as derived by us, will also find its applications.

The paper is organized as follows. In Sec.\ \ref{scalar} we introduce amputed Green function
in CFT through the example of massless scalar field theory. In Sec.\ \ref{spinor}, we introduce
amputation of spinor leg through massless Yukawa field theory. In Sec.\ \ref{SSV}, we give the
structures $C_{1\mu}$ and $C_{2\mu}$ of 
the conformal spinor-spinor-vector Green function and state how spinor leg amputation for these structures
turns out to be different from the Yukawa case. The 
star-triangle relation with two spinors and one vector field is derived in Sec.\ \ref{star}.
Spinor leg amputation of $C_{1\mu}$ and $C_{2\mu}$ is carried out in Secs.\ \ref{C1amp}
and \ref{C2amp}. 
Vector leg amputation
of $C_{1\mu}$ and $C_{2\mu}$ is carried out in
Sec.\ \ref{vecamp}. 
In Sec.\ \ref{current}, checks on these results are performed.
In Sec.\ \ref{concl}, we present our conclusions.
\section{Amputation in scalar field theory}
\label{scalar}
In this section, we explain the aim of our work by reviewing the simplest example of 
scalar field theory.
The two-point function and its inverse for a conformal scalar of scale dimension $d$
are given by
\bea
G_d(x_{12})= \frac{N}{(x_{12}^2)^d}\,,~~~ G_d^{-1}(x_{12})=\frac{\pi^{-D}}{N}\,
             \frac{\Gamma(d)\Gamma(D-d)}{\Gamma(D/2-d)\Gamma(d-D/2)}\,
             \frac {1}{(x_{12}^2)^{D-d}}                          \label{Ginv}
\eea
where $x_{ab}\equiv x_a-x_b$ and $N$ is an arbitrary constant. Together they satisfy
\bea
\int d^Dx_2~ G_d(x_{12})G_d^{-1}(x_{23})=\int d^Dx_2~ G_d^{-1}(x_{12}) G_d(x_{23})=\delta^{(D)}(x_{13})\,.
                                                                         \label{eq:1}
\eea
Eq.\ (\ref{Ginv}) shows that $G_d^{-1}$ is the two-point function of a scalar field of scale dimension $D-d$.
In Appendix \ref{app:inverse}, we indicate how to arrive at $G_d^{-1}$ from $G_d$.
A field of the
same spin but of scale dimension $D-d$ is called the {\it conformal partner} \cite{book} or 
{\it shadow operator} \cite{parisi1} of the field of scale dimension $d$.
Both these fields have the same set of values for the Casimir operators of the conformal group.

Consider next a three-point function
$\langle \phi_d(x_1)\phi_l(x_2)\phi_\Delta(x_3)\rangle$ of three scalar fields of scale dimensions
$d$, $l$ and $\Delta$. The three-point function with the $\phi_d$-leg amputated is defined by
\bea
\langle \phi_d(x_1)\phi_l(x_2)\phi_\Delta(x_3)\rangle\equiv
\int d^Dy~G_d(x_1-y)\langle \tilde\phi_d(y)\phi_l(x_2)\phi_\Delta(x_3)\rangle\,,
\eea
with $\tilde{}$ on a field denoting amputation.
Using Eq.\ (\ref{eq:1}), this definition can also be written as
\bea
\int d^Dx_1~G_d^{-1}(x_{41})\langle \phi_d(x_1)\phi_l(x_2)\phi_\Delta(x_3)\rangle\equiv
\langle \tilde\phi_d(x_4)\phi_l(x_2)\phi_\Delta(x_3)\rangle\,.       \label{eq:2}
\eea
Next, using the conformal transformation properties of the left-hand side of Eq.\ (\ref{eq:2}),
it can be shown that \cite{book} {\it the amputed three-point function is again a three-point function
but with $\phi_d$ replaced by its conformal partner}. [In Appendix
\ref{app:gen} of the present work, we demonstrate this result for spinor and vector field.] 
Thus,
\bea
\int d^Dx_1~G_d^{-1}(x_{41})\langle \phi_d(x_1)\phi_l(x_2)\phi_\Delta(x_3)\rangle\sim
\langle \phi_{D-d}(x_4)\phi_l(x_2)\phi_\Delta(x_3)\rangle\,,       \label{eq:3}
\eea
where $\sim$ means upto some coefficient. Now, the structure of 
$\langle \phi_d(x_1)\phi_l(x_2)\phi_\Delta(x_3)\rangle$ is known in CFT: it is given by
\bea
C^{d,l,\Delta}(x_1x_2x_3)=\frac{1}{x_{12}^{d+l-\Delta}x_{13}^{d+\Delta-l}x_{23}^{l+\Delta-d}}\,.
\eea
The non-trivial part in determining the coefficient on the right-hand side of Eq.\ (\ref{eq:3}) 
is therefore the evaluation of the integral $\int d^Dx_1~(x_{14}^2)^{-(D-d)}C^{d,l,\Delta}(x_1x_2x_3)$
occuring on the left-hand side.
This can be done by using the star-triangle relation of Eq.\ (\ref{st}). We then find that
\bea 
\int d^Dx_1~G_d^{-1}(x_{41})\,C^{d,l,\Delta}(x_1x_2x_3)&=&
              \frac{\pi^{-D/2}}{N}\,\frac{\Gamma(d)\Gamma(\frac{D-d-l+\Delta}{2})
                             \Gamma(\frac{D-d-\Delta+l}{2})}
                       {\Gamma(\frac{D}{2}-d)\Gamma(\frac{d+l-\Delta}{2})\Gamma(\frac{d+\Delta-l}{2})}
                        \nonumber\\&&\times C^{D-d,l,\Delta}(x_4x_2x_3)\,,            \label{eq:4}
\eea
where $G_d^{-1}$ is given in Eq.\ (\ref{Ginv}). Thus the result given in Eq. (\ref{eq:3}) is explicitly 
realized in Eq. (\ref{eq:4}).
[Let us note that that Eq.\ (2.11) of Ref.\ \cite{georgi'} can be reproduced from our 
Eq.\ (\ref{eq:4})
by relabelling the scale dimensions and the coordinates appropriately.]
{\it The aim of the present work is to derive similar
amputation equations for the spinor-spinor-vector Green function} which is relevant to QED.
\section{Amputation of spinor leg}
\label{spinor}
The fermion two-point function $\langle\psi_d(x_1)\bar\psi_d(x_2)$ and its inverse 
in CFT are given by
\bea
S_d(x_{12})&=&N\frac{\rlap/x_{12}}{(x_{12}^2)^{d+1/2}}\,,\nonumber\\
S_d^{-1}(x_{12})&=&-\frac{\pi^{-D}}{N}\,
                \frac{\Gamma(d+1/2)\Gamma(D-d+1/2)}{\Gamma(D/2-d+1/2)\Gamma(d-D/2+1/2)}\,
                    \frac{\rlap/x_{12}}{(x_{12}^2)^{D-d+1/2}}                 \label{Sinv}
\eea
which satisfy Eq.\ (\ref{eq:1}) with $G_d$ replaced by $S_d$ (see Appendix \ref{app:inverse}).
Here $N$ is again an arbitrary constant.
It will be instructive to first consider the Yukawa ($\bar\psi\gamma_5\psi\phi$) theory
($D$ is even and $\gamma_5=i^{D/2}\gamma_1\gamma_2\cdots\gamma_D$).
There are two conformal-invariant structures \cite{book} for 
$\langle \psi_d(x_1)\bar\psi_l(x_2)\phi_\Delta(x_3)\rangle$:
\bea
C_+^{d,l,\Delta}(x_1x_2x_3)&=&\frac{\rlap/x_{13}}{x_{13}^{d-l+\Delta+1}}\,\gamma_5\,
                            \frac{\rlap/x_{32}}{x_{23}^{l-d+\Delta+1}}\,
                            \frac{1}{x_{12}^{l+d-\Delta}}\,,               \label{C+}\\
C_-^{d,l,\Delta}(x_1x_2x_3)&=&\frac{\rlap/x_{12}}{x_{12}^{l+d-\Delta+1}}\,\gamma_5\,
                            \frac{1}{x_{13}^{d-l+\Delta}}\,
                            \frac{1}{x_{23}^{l-d+\Delta}}\,                  \label{C-}
\eea
with $\gamma_5 C_\pm \gamma_5=\pm C_\pm$.  
Corresponding to Eq.\ (\ref{eq:4}), we now have
\bea
\int d^Dx_1~S_d^{-1}(x_{41})\,C_\pm^{d,l,\Delta}(x_1x_2x_3)
&=&K_\pm C_\mp^{D-d,l,\Delta}(x_4x_2x_3)             \label{yu1}\\
\int d^Dx_2~C_\pm^{d,l,\Delta}(x_1x_2x_3)S_l^{-1}(x_{25})
&=&K'_\pm C_\mp^{d,D-l,\Delta}(x_1x_5x_3)              \label{yu2}
\eea
the integrals being evaluated by using the star-triangle relation for 
the Yukawa theory given in Eq.\ (\ref{st'}). Here $S_d^{-1}$ and $S_l^{-1}$ are as  given 
in Eq.\ (\ref{Sinv}), and
\bea
K_+&=&-\frac{\pi^{-D/2}}{N}\,\frac{\Gamma(d+\frac{1}{2})\Gamma(\frac{D-d+l-\Delta+1}{2})
                             \Gamma(\frac{D-d-l+\Delta}{2})}
	{\Gamma(\frac{D}{2}-d+\frac{1}{2})
	  \Gamma(\frac{d-l+\Delta+1}{2})\Gamma(\frac{d+l-\Delta}{2})}\,,\nonumber\\
K_-&=&\frac{\pi^{-D/2}}{N}\,\frac{\Gamma(d+\frac{1}{2})\Gamma(\frac{D-d+l-\Delta}{2})
                             \Gamma(\frac{D-d-l+\Delta+1}{2})}
	{\Gamma(\frac{D}{2}-d+\frac{1}{2})
	    \Gamma(\frac{d-l+\Delta}{2})\Gamma(\frac{d+l-\Delta+1}{2})}\,,\nonumber\\
K'_+&=&-K_+|_{d\leftrightarrow l}\,,~~~K'_-=-K_-|_{d\leftrightarrow l}\,.
\eea
For the case $D=4$, these results are given in a different form in Appendix 6 of Ref.\
\cite{fradkin1}. It may be noted that
amputation again replaces $d$ by $D-d$ (or $l$ by $D-l$) in Eqs.\ (\ref{yu1}) and (\ref{yu2}) 
in accordance with the
general result. An additional feature is that $C_+$ goes over to $C_-$ and vice versa
in these equations. This is consistent with the counting of the number of gamma
matrices on each side of Eq.\ (\ref{yu1}) and Eq.\ (\ref{yu2}). The point is that we
must have either odd or even number of gamma matrices on each side of an equation
(since the product of odd (even) number of gamma matrices has a zero (non-zero) trace for even $D$).
That amputation of {\it one} spinor leg gives back a standard structure is a special feature of the
Yukawa theory. We will see that this feature is not present when we have a vector field
coupling to the spinors.
\section{Spinor-spinor-vector Green function}
\label{SSV}
The Green function 
$\langle \psi_d(x_1)\bar\psi_l(x_2)\Phi_\mu^\Delta(x_3)\rangle$ has 
two conformal invariant structures \cite{book}:
\bea
C_{1\mu}^{d,l,\Delta}(x_1x_2x_3)&=&\frac{\rlap/x_{13}\gamma_\mu\rlap/x_{32}}
                       {x_{12}^{d+l-\Delta}x_{13}^{d-l+\Delta+1}x_{23}^{l-d+\Delta+1}}\,,
                                                              \label{C1mu}\\
C_{2\mu}^{d,l,\Delta}(x_1x_2x_3)&=&\frac{\rlap/x_{12}}{x_{12}^{d+l-\Delta+2}}\,
                           \Bigg(\frac{x_{13\mu}}{x_{13}^{d-l+\Delta+1}x_{23}^{l-d+\Delta-1}}-
                                 \frac{x_{23\mu}}{x_{13}^{d-l+\Delta-1}x_{23}^{l-d+\Delta+1}}\Bigg)
                                                                \label{C2mu}\\
                                 &=&\frac{\rlap/x_{12}}
                       {x_{12}^{d+l-\Delta+2}x_{13}^{d-l+\Delta-1}x_{23}^{l-d+\Delta-1}}\,
                       \lambda_\mu^{x_3}(x_1x_2)\,,
\eea
where
\bea
\lambda_\mu^{x_3}(x_1x_2)=\frac{x_{13\mu}}{x_{13}^2}-\frac{x_{23\mu}}{x_{23}^2}\,.
\eea
[These structures are also given in Refs.\ \cite{fradkin2}, \cite{palchik} and \cite{comm}, but only
for the case $\Delta=1$.]

To ampute $\psi_d$ (say), we have to proceed as in Eq. (\ref{yu1}). But we will now come across
an important difference:
{\it the amputation of one spinor leg
will not give back either $C_{1\mu}$ or $C_{2\mu}$ (or a linear combination
of them).} At least for even $D$,
this can be understood from the fact that
each $\int d^Dx_1~(\rlap/x_{41}/(x_{14}^2)^{D-d+1/2})
C_{i\mu}$ (with $i=1,2$) is a product of even number of gamma matrices, while both
(\ref{C1mu}) and (\ref{C2mu}) have odd number of gamma matrices. 
In order to be consistent
with the general result, the structures resulting from amputing $\psi_d$
will still be conformal invariant with the
expected values of scale dimensions (i.e. $D-d$, $l$ and $\Delta$).
[We explicitly check the conformal invariance of such structures are in Appendix \ref{app:non}.]
But these structures are non-standard
in the sense that they do not have any symmetry under the interchange of the two fermions
and hermitian conjugation. [On the other hand, the standard structures in Eqs.\ (\ref{C1mu})
and (\ref{C2mu}), and also those in Eqs.\ (\ref{C+}) and (\ref{C-}), are invariant
when $x_1\leftrightarrow x_2$, $d\leftrightarrow l$ and hermitian conjugation
are performed together. Recall that the Euclidean gamma matrices are all hermitian.] However,
when $\it both$ $\psi_d$ and $\psi_l$ are amputated, we get back linear combinations
of $C_{1\mu}$ and $C_{2\mu}$: see Secs.\ \ref{C1amp} and \ref{C2amp}.
\section{Star-triangle relation with two spinors and one vector field}
\label{star}
The star-triangle relation which we are going to prove, and which will be later used for amputing 
$C_{1\mu}^{d,l,\Delta}$, is:
\bea
&&\int d^Dx_4~ \frac{\rlap/x_{14}}{(x_{14}^2)^{\delta_1+1/2}}\,\gamma_\nu
              \frac{\rlap/x_{42}}{(x_{24}^2)^{\delta_2+1/2}}
               \frac{g_{\mu\nu}(x_{34})}{(x_{34}^2)^{\delta_3}}\nonumber\\
&=&\pi^{D/2}\frac{ \Gamma(D/2-\delta_1+1/2) \Gamma(D/2-\delta_2+1/2) \Gamma(D/2-\delta_3)}
{\Gamma(\delta_1+1/2)\Gamma(\delta_2+1/2)\Gamma(\delta_3+1)}\nonumber\\
&&\times\Bigg((\delta_3-1) \frac{\rlap/x_{13}\gamma_\mu\rlap/x_{32}}
                 {(x_{12}^2)^{D/2-\delta_3}(x_{13}^2)^{D/2-\delta_2+1/2}
                  (x_{23}^2)^{D/2-\delta_1+1/2}}\nonumber\\
          &&+(D-2\delta_3)\,\frac{\rlap/x_{12}}
     {(x_{12}^2)^{D/2-\delta_3+1}(x_{13}^2)^{D/2-\delta_2-1/2}
                  (x_{23}^2)^{D/2-\delta_1-1/2}}\lambda_\mu^{x_3}(x_1x_2)\Bigg)
                                                                                \label{st''}
\eea
where Eq. (\ref{delta}) holds. The vector field has the propagator corresponding to scale
dimension $\delta_3$ (see Eq.\ (\ref{Dinv})) with
\bea
g_{\mu\nu}(x)=\delta_{\mu\nu}-2\frac{x_\mu x_\nu}{x^2}\,.              \label{gmunu}
\eea
Eq.\ (\ref{st''}) can be viewed as  {\it analogous to the more familiar star-triangle
relations} given by
Eqs.\ (\ref{st}) and (\ref{st'}), as follows.
The left-hand side of Eq.\ (\ref{st''}) represents the propagation of two conformal
spinors and one conformal vector field from the external points $x_a$ ($a=1,2,3$)
to the internal vertex $x_4$ with an interaction $\gamma_\nu$. The right-hand side
is a linear combination of the two available structures (\ref{C1mu}) and (\ref{C2mu}).

A check for Eq.\ (\ref{st''}) can be performed for the case $\delta_3=1$. In this case,
the vector field propagator on the left-hand side is 
$g_{\mu\nu}(x_{34})/x_{34}^2=\partial^\mu_{x_3}
\partial^\nu_{x_3}\ln|x_{34}|$, that is, longitudinal in $x_3$.
On the right-hand side, only the second term remains, and this term
is also longitudinal in $x_3$ as follows. Since Eq.\ (\ref{delta}) now gives
$D/2-\delta_1-1/2=-(D/2-\delta_2-1/2)=n/2$ (say), the coordinate $x_3$ now occurs in
the combination $(x_{13}/x_{23})^n\lambda_\mu^{x_3}(x_1x_2)$, which equals
$-(1/n)\partial^\mu_{x_3}(x_{13}/x_{23})^n$.

A relation previously derived in
Refs.\ \cite{my1} and \cite{my2} also involved two spinors and one vector field.
But {\it there are two important differences between that relation and the
relation (\ref{st''}) above.} Firstly, the previous relation had a covariant gauge propagator,
while the vector field propagator in Eq.\ (\ref{st''}) is invariant under the standard
transformation law for a conformal vector. This is necessary for the amputation of the spinor leg 
and also the vector leg in the structure
$C_{1\mu}^{d,l,\Delta}$ \cite{footnote1}. 
The other difference is that here we have completely general
values for the scale dimensions $\delta_1$, $\delta_2$ and $\delta_3$; 
this will also be necessary for the present purpose.

{\it The derivation of Eq.\ (\ref{st''})}, which is to be presented now, {\it will be
along the same lines as followed in Refs.\ \cite{my1} and \cite{my2}}.
We are thus going to use the operator algebraic method due to Isaev \cite{isaev1}
which reduces Feynman integrals to products
of position and momentum operators $\hat{q}_i$ and $\hat{p}_i$ taken between
position eigenstates.
As explained in Sec.\ 2 of Ref.\ \cite{my1}, this method involves starting from the
$``\hat{p}\hat{q}\hat{p}"$ form and passing to the
$``\hat{q}\hat{p}\hat{q}"$ form, using the key relation given by Eq.\ (\ref{key1}). In our case, the idea is to split the left-hand side
of Eq.\ (\ref{st''}) into a longitudinal part and a transverse part, and tackle them
as in Sec.\ 4 of Ref.\ \cite{my1} and Sec.\ 2 of Ref.\ \cite{my2} respectively. In view
of the general values of the scale dimensions, the starting $``\hat{p}\hat{q}\hat{p}"$ forms
are somewhat different from that in these references. The starting forms are
\bea
\Gamma^{\rm long}_\mu&\equiv&
(\gamma_\lambda\gamma_\nu\gamma_\rho \hat{p}_\lambda\hat{p}^{\,-2\alpha-1} \hat{q}_\rho
\hat{q}^{\,-2(\alpha+\beta)-1}\hat{p}^{\,-2\beta})\hat{p}_\nu\hat{p}_\mu\hat{p}^{\,-2}\,,
                                                                      \label{Gl}\\
\Gamma^{\rm tr}_\mu&\equiv&
(\gamma_\lambda\gamma_\nu\gamma_\rho \hat{p}_\lambda\hat{p}^{\,-2\alpha-1}
\hat{q}_\rho
\hat{q}^{\,-2(\alpha+\beta)-1}\hat{p}^{\,-2\beta})
(\delta_{\nu\mu}-\hat{p}_\nu\hat{p}_\mu\hat{p}^{\,-2})\,.                \label{Gt}
\eea
[These are, however, quite similar to the $``\hat{p}\hat{q}\hat{p}"$ form for the
three-point function of the Yukawa theory: see Eq.\ (5) of Ref.\ \cite{my1}.] 
{\it A new element in this calculation is that 
$\Gamma^{\rm long}_\mu$ and $\Gamma^{\rm tr}_\mu$ have to be taken
in a precise proportion} to result in the structure (\ref{gmunu}). To determine this proportion,
we use the relation
\bea
\frac{g_{\mu\nu}(x)}{r^n}=\frac{n-2}{n}\Bigg(\Big(\delta_{\mu\nu}-
                     \frac{\partial_\mu\partial_\nu}{\partial^2}\Big)
                    +\frac{2D-n-2}{n-2}\,\frac{\partial_\mu\partial_\nu}{\partial^2}\Bigg)
                    \frac{1}{r^n}\,.                                       \label{gg}
\eea
[This formula can be derived by first evaluating $\partial_\mu\partial_\nu(1/r^{n-2})$, and
hence $\partial^2(1/r^{n-2})$. Here $r=\sqrt{x_\mu x_\mu}$.] Since $\hat{p}^{\,-2\beta}$
in Eqs.\ (\ref{Gl}) and (\ref{Gt}) goes as $r^{-(D-2\beta)}$ in position space, we need to
consider the case $n=D-2\beta$ in Eq.\ (\ref{gg}). Thus we have to start with the
$``\hat{p}\hat{q}\hat{p}"$ form
\bea
\Gamma_\mu=\Gamma^{\rm tr}_\mu+\frac{D+2\beta-2}{D-2\beta-2}\Gamma^{\rm long}_\mu\,.
                                                                                \label{Gmu}
\eea

Next, we have to express the position-space matrix elements (that is,
between $\langle x|$ and $|y\rangle$)
of the right-hand sides of Eqs.\ (\ref{Gl}) and 
(\ref{Gt}) in terms of the matrix elements of 
$\hat{p}_\lambda\hat{p}^{\,-2\alpha-1}$,
$\hat{p}^{\,-2\beta}$ and $\hat{p}_\nu\hat{p}_\mu\hat{p}^{\,-2}$ from the Appendix of
Ref.\ \cite{my1} (see Eqs.\ (14) and (15) of Ref.\ \cite{my1}). Then we can write down the
matrix element of the right-hand side of Eq.\ (\ref{Gmu}) using Eq.\ (\ref{gg})
for $n=D-2\beta$. This leads to
\bea
\langle x|\Gamma_\mu|y\rangle&=&\frac{i\Gamma(D/2-\alpha+1/2)\Gamma(D/2-\beta+1)}
                                   {\pi^D 2^{2\alpha+2\beta-1}(D-2\beta-2)\Gamma(\alpha+1/2)\Gamma(\beta)}
\nonumber\\
&&\int d^Dz~\frac{\rlap/x-\rlap/z}{|x-z|^{D-2\alpha+1}}\,\gamma_\nu
            \frac{\rlap/z}{|z|^{2(\alpha+\beta)+1}}\,
            \frac{g_{\mu\nu}(y-z)}{|y-z|^{D-2\beta}}\,.               \label{star1} 
\eea

On the other hand, we can put $\Gamma_\mu$ in the $``\hat{q}\hat{p}\hat{q}"$ form
and then take the matrix element. This involves a long calculation, given in
Appendix \ref{app:steps} (the important points are highlighted after Eqs.\ (\ref{app6}) and (\ref{relation}).)
It leads to
\bea
\langle x|\Gamma_\mu|y\rangle=\frac{i\Gamma(D/2-\alpha-\beta+1/2)}
                                   {\pi^{D/2} 2^{2\alpha+2\beta}\Gamma(\alpha+\beta+1/2)}\,
\frac{x^2(\rlap/x-\rlap/y)\gamma_\mu\rlap/y+\frac{4\beta}{D-2\beta-2}
                            ((x-y)^2y_\mu+y^2(x-y)_\mu)\rlap/x}
         {x^{2\beta+2}|x-y|^{D-2\alpha-2\beta+1}y^{2\alpha+1}}\,.      \label{star2}
\eea
The right-hand sides of Eqs.\ (\ref{star1}) and (\ref{star2}) are now to be equated.
After that, we let
$x=x_1-x_2$ and  $y=x_3-x_2$, and also change to a new integration
variable $x_4$ defined by $z=x_4-x_2$. We also define $\delta_1$, $\delta_2$ and $\delta_3$ by
$D/2-\alpha=\delta_1$,  $\alpha+\beta=\delta_2$ and $D/2-\beta=\delta_3$.
This leads us to the relation given in Eq.\ (\ref{st''}).
\section{Spinor leg amputation in $C_{1\mu}^{d,l,\Delta}(x_1x_2x_3)$}
\label{C1amp}
In this Section and the next, we are going to evaluate
\bea
\int d^Dx_1\,d^Dx_2~\frac{\rlap/x_{41}}{(x_{14}^2)^{D-d+1/2}}C_{i\mu}^{d,l,\Delta}(x_1x_2x_3)
\frac{\rlap/x_{25}}{(x_{25}^2)^{D-l+1/2}}\,,~~~~i=1,2.                   \label{C1eq1}
\eea
The integration over $x_1$ amputes $\psi_d(x_1)$, while that over $x_2$ amputes $\bar\psi_l(x_2)$.

We consider $C_{1\mu}$ in this Section. From Eqs.\ (\ref{C1eq1}) and  (\ref{C1mu}),
we see that the $x_1$ integration can be done by using the star-triangle relation of Eq.\ 
(\ref{st'}). The integration over $x_2$ then involves
\bea
\rlap/x_{23}\gamma_\mu\rlap/x_{32}= x_{23}^2 \gamma_\nu g_{\mu\nu}(x_{23})\,.
\eea
Consequently, this integral is of the form
\bea
\int d^Dx_2~\frac{\rlap/x_{42}}{x_{24}^{D-d+l-\Delta+1}}\gamma_\nu
\frac{\rlap/x_{25}}{(x_{25}^2)^{D-l+1/2}}\,\frac{g_{\mu\nu}(x_{23})}
{x_{23}^{d+l+\Delta-D}}                                                           \label{eq:28}
\eea
which can be evaluated by using the star-triangle relation of Eq.\
(\ref{st''}). We thus get
\bea
&&\int d^Dx_1\,d^Dx_2~S_d^{-1}(x_{41})\,C_{1\mu}^{d,l,\Delta}(x_1x_2x_3)\,S_l^{-1}(x_{25})\nonumber\\
&=&F(d,l,\Delta)\times(d+l-\Delta)
\Bigg(\frac{d+l+\Delta-D-2}{2}C_{1\mu}^{D-d,D-l,\Delta}(x_4x_5x_3)\nonumber\\
&&+(2D-d-l-\Delta)C_{2\mu}^{D-d,D-l,\Delta}(x_4x_5x_3)\Bigg)                \label{CC1}
\eea
where $S_d^{-1}$ and $S_l^{-1}$ are as in Eq.\ (\ref{Sinv}), and the coefficient $F(d,l,\Delta)$ is given by
\bea
F(d,l,\Delta)=\frac{\pi^{-D}}{2N^2}\,
\frac{\Gamma(d+\frac{1}{2})\Gamma(l+\frac{1}{2})
\Gamma(\frac{D-d-l+\Delta}{2})\Gamma(\frac{2D-d-l-\Delta}{2})}
           {\Gamma(\frac{D}{2}-d+\frac{1}{2})\Gamma(\frac{D}{2}-l+\frac{1}{2}) 
	   \Gamma(\frac{d+l-\Delta+2}{2})\Gamma(\frac{d+l+\Delta-D+2}{2})}\,.
	                                                                      \label{eqF}
\eea
\section{Spinor leg amputation in $C_{2\mu}^{d,l,\Delta}(x_1x_2x_3)$}
\label{C2amp}
Using Eq.\ (\ref{C2mu}),
we write down the integral (\ref{C1eq1}) for $C_{2\mu}$ in full.
There are two terms. On interchanging the integration variables $x_1$, $x_2$ in the
second term, we find that the integral under consideration is
\bea
\int d^Dx_1\,d^Dx_2~\frac{\rlap/x_{41}}{(x_{14}^2)^{D-d+1/2}}
\,\frac{\rlap/x_{12}}{x_{12}^{d+l-\Delta+2}}\,\frac{x_{13\mu}}{x_{13}^{d-l+\Delta+1}}\,
                                 \frac{1}{x_{23}^{l-d+\Delta-1}}\,
\frac{\rlap/x_{25}}{(x_{25}^2)^{D-l+1/2}}\nonumber\\
+\Bigg({\rm hermitian~ conjugate},x_4\leftrightarrow x_5, d\leftrightarrow l\Bigg)
                                       \label{C2amp1}
\eea
Let us evaluate the first term in (\ref{C2amp1}). First we perform the $x_2$ integration
using Eq.\ (\ref{st'}). Then the remaining $x_1$ integral is of the form
\bea
&&\int d^Dx_1~\frac{\rlap/x_{41}}{x_{14}^{2(D-d)+1}}
\,\frac{\rlap/x_{13}x_{13\mu}}{x_{13}^{d+l+\Delta-D+2}}\,
\frac{1}{x_{15}^{D+d-l-\Delta+1}}\nonumber\\
&=&\frac{1}{D-d-l-\Delta}\Bigg(\Big(\frac{1}{2(D-d)-1}\frac{\partial}{\partial\rlap/x_4}
\frac{\partial}{\partial x_{3\mu}}I\Big)\rlap/x_{43}
+\frac{\partial}{\partial x_{3\mu}}I\Bigg)\,,                      \label{ccc}
\eea
where
\bea
I&=&\int d^Dx_1~\frac{1}{x_{14}^{2(D-d)-1} x_{13}^{d+l+\Delta-D} x_{15}^{D+d-l-\Delta+1}}\,.
                                                              \label{cI}
\eea
The right-hand side of Eq.\ (\ref{ccc}) is obtained by writing $\rlap/x_{13}=\rlap/x_{43}
-\rlap/x_{41}$ on the left-hand side. Now $I$ can be evaluated by using Eq.\ (\ref{st}).
After some algebra, the first term in (\ref{C2amp1}) is found to be
\bea
\frac{\Big((l-d+\Delta-1)x_{35}^2x_{34\mu}+(2d-D+1)x_{34}^2x_{35\mu}\Big)
\frac{\rlap/x_{45}}{x_{45}^2}+\frac{l-d+\Delta-1}{2D-d-l-\Delta}\rlap/x_{43}\gamma_\mu\rlap/x_{35}}
{x_{34}^{l-d+\Delta+1}x_{35}^{d-l+\Delta+1}x_{45}^{2D-d-l-\Delta}}            
\eea
multiplied with a coefficient which is symmetric in $d$ and $l$. Then adding 
the second term in (\ref{C2amp1}), we finally arrive at
\bea
&&\int d^Dx_1\,d^Dx_2~S_d^{-1}(x_{41})\,C_{2\mu}^{d,l,\Delta}(x_1x_2x_3)\,S_l^{-1}(x_{25})\nonumber\\
&=&F(d,l,\Delta)\times
\Bigg((\Delta-1)C_{1\mu}^{D-d,D-l,\Delta}(x_4x_5x_3)\nonumber\\
&&+\frac{(2D-d-l-\Delta)(d+l-\Delta-D+2)}{2}C_{2\mu}^{D-d,D-l,\Delta}(x_4x_5x_3)\Bigg)\label{CC2}
\eea
where the coefficient $F(d,l,\Delta)$ is given by Eq.\ (\ref{eqF}).

For the special case of $\Delta=1$, the amputation of spinor legs in $C_{2\mu}$ can be done much more
easily. The reason is that in this case the $x_3$-dependent part in $C_{2\mu}$ equals
$(\partial/\partial x_{3\mu})(x_{23}/x_{13})^{d-l}$, and we need to just apply Eq. (\ref{st'}).
\section{Vector leg amputation in $C_{1\mu}^{d,l,\Delta}(x_1x_2x_3)$
and $C_{2\mu}^{d,l,\Delta}(x_1x_2x_3)$}
\label{vecamp}
The vector field two-point function and its inverse are given by
\bea
D_{\mu\nu}(x_{12})&=&N\frac{g_{\mu\nu}(x_{12})}{(x_{12}^2)^\Delta}\,,\nonumber\\
D_{\mu\nu}^{-1}(x_{12})&=&\frac{\pi^{-D}}{N}\,\frac{\Delta}{D-\Delta-1}\,
                \frac{\Gamma(\Delta-1)\Gamma(D-\Delta+1)}
		{\Gamma(\frac{D}{2}-\Delta)\Gamma(\Delta-\frac{D}{2})}\,
                    \frac{g_{\mu\nu}(x_{12})}{(x_{12}^2)^{D-\Delta}}\,.            \label{Dinv}
\eea
They satisfy 
\bea
\int d^Dx_2~ D_{\mu\nu}(x_{12})D_{\nu\rho}^{-1}(x_{23})
=\int d^Dx_2~ D_{\mu\nu}^{-1}(x_{12}) D_{\nu\rho}(x_{23})=\delta_{\mu\rho}\delta^{(D)}(x_{13})\,.
\eea
The amputation equations are
\bea
\int d^Dx_3\,C_{1\mu}^{d,l,\Delta}(x_1x_2x_3)\,D_{\mu\nu}^{-1}(x_{34})
&=&F'(d,l,\Delta)\times\Big((D-\Delta-1)C_{1\nu}^{d,l,D-\Delta}(x_1x_2x_4)\nonumber\\
                     &&+(2\Delta-D)C_{2\nu}^{d,l,D-\Delta}(x_1x_2x_4)\Big)\,,    \label{vecamp1}
\eea
which is obtained by using Eq.\ (\ref{st''}), and
\bea
\int d^Dx_3\,C_{2\mu}^{d,l,\Delta}(x_1x_2x_3)\,D_{\mu\nu}^{-1}(x_{34})
=F'(d,l,\Delta)\times (\Delta-1)C_{2\nu}^{d,l,D-\Delta}(x_1x_2x_4)\,,             \label{vecamp2}
\eea
which is obtained by using Eq.\ (\ref{st'''}).
Here the coefficient $F'(d,l,\Delta)$ is given by
\bea
F'(d,l,\Delta)=\frac{\pi^{-D/2}}{N}\,\frac{\Delta}{D-\Delta-1}\,
               \frac{\Gamma(\Delta-1)\Gamma(\frac{D-d+l-\Delta+1}{2})
                       \Gamma(\frac{D-l+d-\Delta+1}{2})}
              {\Gamma(\frac{D}{2}-\Delta)\Gamma(\frac{d-l+\Delta+1}{2})\Gamma(\frac{l-d+\Delta+1}{2})}\,.
\eea
\section{Checking results by amputation in $C_{1\mu}-C_{2\mu}$}
\label{current}
First consider the case $\Delta=D-1$, which is the scale dimension of the current.
In this case we have the relation 
\bea
\frac{\partial}{\partial x_{3\mu}}\Big(C_{1\mu}^{d,l,D-1}(x_1x_2x_3)-C_{2\mu}^{d,l,D-1}(x_1x_2x_3)\Big)=0\,.   \label{curr}
\eea
[The case $d=l$ of Eq. (\ref{curr}) is consistent with the fact that both $C_{1\mu}$ and $C_{2\mu}$
satisfy
\bea
\frac{\partial}{\partial x_{3\mu}}C_{1\mu}^{d,d,D-1}(x_1x_2x_3)
=\frac{\partial}{\partial x_{3\mu}}C_{2\mu}^{d,d,D-1}(x_1x_2x_3)
=-\frac{2\pi^{D/2}}{\Gamma(D/2)}(\delta(x_{13})-\delta(x_{23}))\frac{\rlap/x_{12}}{x_{12}^{2d+1}}\,,
\eea
which is the Ward identity in position space.]
Now from Eq.\ (\ref{C1eq1}), we see that $\partial/\partial x_{3\mu}$ commutes with the operation
 of spinor leg amputation (also, $\Delta$ stays $D-1$ after this amputation). 
Thus, {\it the combination $C_{1\mu}-C_{2\mu}$ should continue to be of this form
after the spinor legs are amputed.}
Indeed, by putting $\Delta=D-1$ in Eqs.\ (\ref{CC1}) and (\ref{CC2}) and taking
the difference, we obtain
\bea
&&\int d^Dx_1\,d^Dx_2~S_d^{-1}(x_{41})\,\Big(C_{1\mu}^{d,l,D-1}(x_1x_2x_3)
-C_{2\mu}^{d,l,D-1}(x_1x_2x_3)\Big)\,S_l^{-1}(x_{25})                         \nonumber\\
&=&
\frac{1}{2}F(d,l,D-1)\times(d+l-1)(d+l-D-1)             \nonumber\\
&&\times\Big(C_{1\mu}^{D-d,D-l,D-1}(x_4x_5x_3)-C_{2\mu}^{D-d,D-l,D-1}(x_4x_5x_3)\Big)
\eea
This serves as a check on the coefficients obtained in Eqs.\ (\ref{CC1}) and (\ref{CC2}).

Next, for any $d$, $l$ and $\Delta$, the two structures satisfy the relation
\bea
{\rm Tr}\Big[\rlap/x_{12}\Big(C_{1\mu}^{d,l,\Delta}(x_1x_2x_3)
-C_{2\mu}^{d,l,\Delta}(x_1x_2x_3)\Big)\Big]=0\,.
\eea
Now vector leg amputation involves only $x_3$. So {\it $C_{1\mu}-C_{2\mu}$ should stay in this
combination after the vector leg is amputed.} From Eqs.\ (\ref{vecamp1}) and (\ref{vecamp2}), we indeed find that
\bea
&&\int d^Dx_3\,\Big(C_{1\mu}^{d,l,\Delta}(x_1x_2x_3)-C_{2\mu}^{d,l,\Delta}(x_1x_2x_3)\Big)\,D_{\mu\nu}^{-1}(x_{34})
\nonumber\\
&=&
F'(d,l,\Delta)\times(D-\Delta-1)\Big(C_{1\nu}^{d,l,D-\Delta}(x_1x_2x_4)
                     -C_{2\nu}^{d,l,D-\Delta}(x_1x_2x_4)\Big)\,,
\eea
which checks the coefficients in the amputation equations.
\section{Conclusion}\label{concl}
We have derived a new star-triangle relation involving two spinors and one vector field, with
general values of the scale dimensions. The relation has been applied to amputation of conformal
invariant three-point function involving these fields. The star-triangle relation
can be of general use in conformal field theoretical context.

Amputed Green functions are the coefficients of the various quasi-primary fields in the conformal
partial wave expansion of the product of two field operators, and thereby enter the partial wave
expansion of the four-point function.
Two examples of application of amputation equations
in these context are the solution of the Thirring model in Ref.\ \cite{fradkin1} and the calculation
of various unparticle processes in the Sommerfield model in Ref.\ \cite{georgi'}.
We have extended the existing work on amputation of conformal Green functions to the next complicated case
involving two spinors and a vector field, 
With the infrared limit of massless QED$_3$ as one of the possible areas of application.

A general discussion of conformal partial wave expansion for the case of more than one independent
invariant structure is given in Ref. \cite{fradkin2}. (That discussion does not deal with spinor legs,
but the presence of more than one structure applies to our work also.) In this case, the proper choice of
linear combination of the independent structures is determined by the orthogonality condition, which
involves amputed Green functions \cite{footnote2}. The partial wave expansion then takes a diagonal form.
Amputation equations will thus be necessary for finding explicit realization of such a situation. 

\section*{Acknowledgement}
The author completed writing up the first version of this work during a visit to IMSc, Chennai. He thanks IMSc
for kind hospitality and H.\ S.\ Sharatchandra for useful discussion during the visit.

\appendix
\leftline{\null\hrulefill\null}\nopagebreak
\section*{Appendices}
\section{Inverse of two-point function}\label{app:inverse}
The inverses of the two-point functions are given in Eqs.\ (\ref{Ginv}), (\ref{Sinv}) and 
(\ref{Dinv}). Once we settle for a value of $N$ in any of these equations, the coefficient
of the inverse two-point function gets fixed by an integral like
$\int d^Dx_2~ x_{12}^{-2d}x_{23}^{-2(D-d)}$ (this is for the scalar field). In this Appendix, we
indicate how such integrals can be evaluated.

For the scalar field, we have
\bea
\int d^Dx_2~ \frac{1}{x_{12}^{2d}}\,\frac{1}{x_{23}^{2(D-d)}}
=\int d^Dx_2~\frac{\langle x_1|\hat p^{-2(D/2-d)}|x_2\rangle} {a(D/2-d)}\,
             \frac{\langle x_2|\hat p^{-2(d-D/2)}|x_3\rangle} {a(d-D/2)}\,,         \label{appin1}
\eea
using $\langle x|\hat p^{-2\alpha}|y\rangle=a(\alpha)|x-y|^{-(D-2\alpha)}$, with $a(\alpha)$
given in the Appendix of Ref.\ \cite{my1}. Using $\int d^Dx_2~|x_2\rangle\langle x_2|=1$, we
can then evaluate the right-hand side of Eq.\ (\ref{appin1}).

For the spinor field, we can similarly evaluate $\int d^Dx_2~ (\rlap/x_{12}/x_{12}^{2d+1})
(\rlap/x_{23}/x_{23}^{2(D-d)+1})$ by using the expression for 
$\langle x|\hat p_i\hat p^{-2\alpha}|y\rangle$ given in the Appendix of Ref.\ \cite{my1}.
Finally, for the vector field we need to evaluate the integral 
$\int d^Dx_2~(g_{\mu\nu}(x_{12})/x_{12}^{2\Delta})(g_{\nu\rho}(x_{23})/x_{23}^{2(D-\Delta)})$.
Here the method is to  use Eq.\ (\ref{gg}), and convert this integral to a differential operator
acting on the integral given in Eq.\ (\ref{appin1}).
\section{General treatment of amputation of spinor leg and vector leg}\label{app:gen}
Here we demonstrate the result that amputation replaces a spinor or vector field by its its conformal partner in a Green function
using the specific case of the three-point function.
This extends the demonstration for scalar field
given in Ref.\ \cite{book}. 

For amputation of the {\it spinor leg}, we want to show that (compare with Eq.\ (\ref{eq:2}) for the scalar field)
\bea
\langle \tilde\psi_d(x_4)\bar\psi_l(x_2)\Phi_\mu^\Delta(x_3)\rangle\equiv
\int d^Dx_1~S_d^{-1}(x_4-x_1)\langle \psi_d(x_1)\bar\psi_l(x_2)\Phi_\mu^\Delta(x_3)\rangle
\label{gen1}
\eea
is a three-point function with dimensions $D-d$, $l$ and $\Delta$ for the three fields.
For this, we need to check that it satisfies the invariance condition for the three-point function
with these scale dimensions under conformal inversion and under scale transformation.

Let us consider first conformal inversion: $x_\mu\rightarrow Rx_\mu=x_\mu/x^2$.
Under this operation, the various (Euclidean) fields transform as \cite{book}
\bea
&&\psi'_d(x)=\frac{\rlap/x}{(x^2)^{d+1/2}}\,\psi_d(Rx),~~~~
\bar{\psi'}_l(x)=\bar{\psi}_l(Rx)\,\frac{\rlap/x}{(x^2)^{l+1/2}}\,,\\
&&\Phi'{_\mu^\Delta}(x)=\frac{g_{\mu\nu}(x)}{(x^2)^\Delta}\Phi_\nu^\Delta(Rx)\,.
\eea
So the invariance condition $\langle \psi_d(x_1)\bar\psi_l(x_2)\Phi_\mu^\Delta(x_3)\rangle
=\langle \psi'_d(x_1)\bar{\psi'}_l(x_2)\Phi'{_\mu^\Delta}(x_3)\rangle$
implies that
\bea
\langle \psi_d(x_1)\bar\psi_l(x_2)\Phi_\mu^\Delta(x_3)\rangle
=\frac{g_{\mu\nu}(x_3)}{(x_1^2)^{d+1/2}(x_2^2)^{l+1/2}(x_3^2)^\Delta}\,
\rlap/x_1\langle \psi_d(Rx_1)\bar\psi_l(Rx_2)\Phi_\nu^\Delta(Rx_3)\rangle\rlap/x_2\,.
\label{gen2}
\eea
Similarly, the condition $\langle \psi_d(x_1)\bar\psi_d(x_2)\rangle=
\langle \psi'_d(x_1)\bar{\psi'}_d(x_2)\rangle$ implies that
\bea
S_d^{-1}(x_4-x_1)=\frac{1}{(x_1^2 x_4^2)^{D-d+1/2}}\rlap/x_4 S_d^{-1}(Rx_4-Rx_1)\rlap/x_1\,,
\label{gen3}
\eea
since $S_d^{-1}$ is the two-point function of a spinor of dimension $D-d$ (see Eq.\ (\ref{Sinv})).
We insert Eqs.\ (\ref{gen2}) and (\ref{gen3}) on the right-hand side of Eq.\ (\ref{gen1}),
and then let $x_1\rightarrow Rx_1$ (so $x_1^2\rightarrow 1/x_1^2$ and $d^Dx_1\rightarrow
d^Dx_1\,(x_1^2)^{-D}$). Comparing the resulting expression with  
Eq.\ (\ref{gen1}) again, we get
\bea
\langle \tilde\psi_d(x_4)\bar\psi_l(x_2)\Phi_\mu^\Delta(x_3)\rangle
=\frac{g_{\mu\nu}(x_3)}{(x_4^2)^{D-d+1/2}(x_2^2)^{l+1/2}(x_3^2)^\Delta}
\rlap/x_4\langle \tilde\psi_d(Rx_4)\bar\psi_l(Rx_2)\Phi_\nu^\Delta(Rx_3)\rangle\rlap/x_2\,.
\eea
Comparing this with Eq.\ (\ref{gen2}) leads to the desired conclusion.

For the scale transformation $x_\mu\rightarrow \lambda x_\mu$, we proceed along similar lines,
using $\psi'_d(x)=\lambda^d \psi_d(\lambda x)$,
$\bar{\psi'}_l(x)=\lambda^l\bar{\psi}_l(\lambda x)$,
and $\Phi'{_\mu^\Delta}(x)=\lambda^\Delta\Phi{_\mu^\Delta}(\lambda x)$.
Amputation of $\bar\psi_l$ can be handled similarly.

For amputation of the {\it vector leg}, we have to show that 
\bea
\langle \psi_d(x_1)\bar\psi_l(x_2)\tilde\Phi_\nu^\Delta(x_4)\rangle\equiv
\int d^Dx_3~\langle \psi_d(x_1)\bar\psi_l(x_2)\Phi_\mu^\Delta(x_3)\rangle
D^{-1}_{\mu\nu}(x_{34})
\label{gen4}
\eea
is a three-point function with dimensions $d$, $l$ and $D-\Delta$ for the three fields.
The condition $\langle \Phi_\mu^\Delta(x_1)\Phi_\nu^\Delta(x_2)\rangle=
\langle\Phi'{_\mu^\Delta}(x_1)\Phi'{_\nu^\Delta}(x_2)\rangle$ implies that
\bea
D^{-1}_{\mu\nu}(x_{34})=\frac{g_{\mu\rho}(x_3)g_{\nu\sigma}(x_4)}{(x_3^2 x_4^2)^{D-\Delta}}
                       \,D^{-1}_{\rho\sigma}(Rx_{34})                \label{gen5}
\eea
since $D^{-1}_{\mu\nu}$ is the two-point function of a vector field of dimension $D-\Delta$.
We insert Eqs.\ (\ref{gen2}) and (\ref{gen5}) in the right-hand side of Eq.\ (\ref{gen4}),
then let $x_3\rightarrow Rx_3$ and follow the procedure adopted for $\psi_d$.
\section{Conformal invariance of structures obtained by amputing one spinor leg}\label{app:non}
In this Appendix, we directly check that the structures obtained by amputing {\it one}
 spinor leg in $C_{1\mu}$
are indeed conformal invariant, albeit non-standard, structures.
Amputing {\it only} 
$\psi_d(x_1)$ in $C_{1\mu}^{d,l,\Delta}(x_1x_2x_3)$ by using $S_d^{-1}(x_{14})$
(see Sec.\ \ref{C1amp}), we obtain the structure $(\rlap/x_{42}/x_{24}^{D-d+l-\Delta+1})
\gamma_\nu(g_{\mu\nu}(x_{23})/x_{23}^{d+l+\Delta-D})(1/x_{34}^{D-d-l+\Delta})$.
This should be a conformal invariant structure for 
$\langle \psi_{D-d}(x_4)\bar\psi_l(x_2)\Phi_\mu^\Delta(x_3)\rangle$.
Equivalently,
\bea
\frac{\rlap/x_{12}}{x_{12}^{d+l-\Delta+1}}\,
\gamma_\nu\frac{g_{\mu\nu}(x_{23})}{x_{23}^{l-d+\Delta}}\,\frac{1}{x_{13}^{d-l+\Delta}}
\label{non1}
\eea
should be a conformal invariant structure for 
$\langle \psi_d(x_1)\bar\psi_l(x_2)\Phi_\mu^\Delta(x_3)\rangle$.

Our aim is to check the invariance condition expressed in Eq.\ (\ref{gen2}) for this structure.
This amounts to showing that the expression (\ref{non1}) equals the expression
\bea
\frac{g_{\mu\nu}(x_3)}{x_{12}^{d+l-\Delta+1}x_{23}^{l-d+\Delta}x_{13}^{d-l+\Delta}}\,
\rlap/x_1\Bigg(\frac{\rlap/x_1}{x_1^2}-\frac{\rlap/x_2}{x_2^2}\Bigg)\,
\gamma_\rho g_{\nu\rho}(Rx_{23})\rlap/x_2
\label{non2}
\eea
(where we used $Rx_{12}=x_{12}/(|x_1||x_2|)$). The equality of the two expressions can be shown
by using
$g_{\nu\rho}(Rx_{23})=g_{\nu\lambda}(x_3)g_{\lambda\kappa}(x_{23})
g_{\kappa\rho}(x_2)$, $g_{\mu\nu}(x_3)g_{\nu\lambda}(x_3)=\delta_{\mu\lambda}$,
and $\gamma_\rho g_{\kappa\rho}(x_2)=-\rlap/x_2\gamma_\kappa\rlap/x_2/x_2^2$.

Similarly, by amputing {\it only} $\bar\psi_l$ in $C_{1\mu}^{d,l,\Delta}(x_1x_2x_3)$   
we get another conformal-invariant structure for
$\langle \psi_d(x_1)\bar\psi_l(x_2)\Phi{_\mu^\Delta}(x_3)$, namely,
\bea
\gamma_\nu\frac{g_{\mu\nu}(x_{13})}{x_{13}^{d-l+\Delta}}\,
\frac{\rlap/x_{12}}{x_{12}^{d+l-\Delta+1}}\,\frac{1}{x_{23}^{l-d+\Delta}}\,.
\label{non3}
\eea
The structures (\ref{non1}) and (\ref{non3}), being products of even number of gamma matrices,
are independent of $C_{1\mu}$ and $C_{2\mu}$, which are  products of odd number of gamma matrices.
The former are, however, not invariant under interchange of the two fermions and simultaneous 
hermitian conjugation.
But they are valid structures if the two fermions in the Green function are not identical.

By amputing one spinor leg in $C_{2\mu}$ also, we get non-standard structures,
which are more complicated than the two structures (\ref{non1}) and (\ref{non3}).
\section{Some steps in the derivation of the star-triangle relation of Sec.\ \ref{star}}
\label{app:steps}
Here we show how to express $\Gamma_\mu$ as given by Eqs.\ (\ref{Gl}),
(\ref{Gt}) and (\ref{Gmu}) in the $``\hat{q}\hat{p}\hat{q}"$ form, and arrive
at Eq.\ (\ref{star2}). For $\Gamma^{\rm long}_\mu$, we follow Sec.\ 4 of Ref.\ \cite{my1}.
Thus, we write
\bea
\langle x|\Gamma^{\rm long}_\mu|y\rangle=-i\frac{\partial{^y_\mu}}{(\partial^2)^y}\langle x|\Gamma'|y\rangle\,,
                                                  \label{app1}\\
\Gamma'=\gamma_\lambda\gamma_\nu\gamma_\rho \hat{p}_\lambda\hat{p}^{\,-2\alpha-1} \hat{q}_\rho
\hat{q}^{\,-2(\alpha+\beta)-1}\hat{p}^{\,-2\beta}\hat{p}_\nu\,.                   \label{app2}
\eea
[In $\langle x|\Gamma^{\rm long}_\mu|y\rangle$, we insert $\int d^Dz\,|z\rangle\langle z|$
just after $\hat{p}_\nu$. Now $\langle z|\hat{p}_\mu\hat{p}^{\,-2}|y\rangle=-i\partial{^y_\mu}/(\partial^2)^y\delta^{(D)}(y-z)$,
and $\partial{^y_\mu}/(\partial^2)^y$ can be taken outside the integral over $z$.]

In going from $``\hat{p}\hat{q}\hat{p}"$ to the
$``\hat{q}\hat{p}\hat{q}"$ form, the essential idea is to move $\hat{q}_\mu$ (or $\hat{p}_\mu$)
through powers of $\hat{p}^{\,2}$ (or $\hat{q}^{\,2}$) by using
$[\hat{q}_\mu,\hat{p}^{\,2\alpha}]=i2\alpha\hat{p}^{\,2\alpha-2}\hat{p}_\mu$
(or $[\hat{p}_\mu,\hat{q}^{\,2\alpha}]=-i2\alpha\hat{q}^{\,2\alpha-2}\hat{q}_\mu$), so that
one can use the key relation
\bea
\hat{p}^{\,-2\alpha}\hat{q}^{\,-2(\alpha+\beta)}\hat{p}^{\,-2\beta}
=\hat{q}^{\,-2\beta}\hat{p}^{\,-2(\alpha+\beta)}\hat{q}^{\,-2\alpha}\,,              \label{key1}
\eea
at an intermediate stage. Eq.\ (\ref{key1}) is the star-triangle relation of Eq.\ (\ref{st}) in
the operator form. We thus follow the steps in Eqs.\ (19)-(23) of Ref.\ \cite{my1}. In the
present case, this leads to
\bea
\Gamma'&=&\gamma_\rho \hat{q}^{\,-2\beta}\hat{p}^{\,-2(\alpha+\beta)+1}
           \hat{q}^{\,-2\alpha-1}\hat{q}_\rho
       +i2\beta\gamma_\lambda\gamma_\nu\gamma_\rho\hat{q}^{\,-2\beta-2}\hat{q}_\lambda
        \hat{p}^{\,-2(\alpha+\beta)-1}\hat{p}_\nu\hat{q}^{\,-2\alpha-1}\hat{q}_\rho    \nonumber\\
       &&+i(D-2\alpha-2\beta-1) \gamma_\lambda\hat{q}^{\,-2\beta}\hat{p}^{\,-2(\alpha+\beta)-1}
        \hat{p}_\lambda\hat{q}^{\,-2\alpha-1} \nonumber\\
       &&-2\beta(D-2\alpha-2\beta-1)\gamma_\lambda\hat{q}^{\,-2\beta-2}\hat{q}_\lambda
       \hat{p}^{\,-2(\alpha+\beta)-1}
           \hat{q}^{\,-2\alpha-1}\,.                                       \label{four}
\eea
Then we use the various position space matrix elements listed in the Appendix of Ref.\ \cite{my1}
to arrive at
\bea
\langle x|\Gamma'|y\rangle&=&\frac{\Gamma(D/2-\alpha-\beta+1/2)}
                                   {\pi^{D/2} 2^{2\alpha+2\beta-1}\Gamma(\alpha+\beta+1/2)}
                  \nonumber\\
&&\times\frac{(D/2-\beta-1)x^2\rlap/y-(D/2-\alpha-1/2)x^2\rlap/x+2\beta x\cdot y \rlap/x}
         {x^{2\beta+2}|x-y|^{D-2\alpha-2\beta+1}y^{2\alpha+1}}\,.            \label{app4}
\eea
For $\Gamma^{\rm tr}_\mu$, given in Eq.\ (\ref{Gt}), we follow Sec.\ 2 of 
Ref.\ \cite{my2}, and first 
split it into two parts:
\bea
\Gamma^{\rm tr}_\mu&=&\Gamma^{\rm tr(1)}_\mu+\Gamma^{\rm tr(2)}\,,\\
\Gamma^{\rm tr(1)}_\mu&=&2\gamma_\lambda\hat{p}_\lambda
 \hat{p}^{\,-2\alpha-1}\hat{q}_\nu\hat{q}^{\,-2(\alpha+\beta)-1}\hat{p}^{\,-2\beta}
{\cal P}_{\mu\nu}\,,\\
\Gamma^{\rm tr(2)}_\mu&=&-\gamma_\lambda\gamma_\rho\gamma_\nu\hat{p}_\lambda
 \hat{p}^{\,-2\alpha-1}\hat{q}_\rho\hat{q}^{\,-2(\alpha+\beta)-1}\hat{p}^{\,-2\beta}
{\cal P}_{\mu\nu}
\eea
with ${\cal P}_{\mu\nu}=\delta_{\mu\nu}-\hat{p}_\mu\hat{p}_\nu\hat{p}^{\,-2}$.
Following the steps in Eqs.\ (6)-(12) of Ref.\ \cite{my2}, we have
\bea
\Gamma^{\rm tr(1)}_\mu&=&2\gamma_\lambda(\hat{q}^{\,-2\beta}
+i2\beta\hat{q}_\lambda\hat{q}^{\,-2\beta-2})
\hat{p}^{\,-2(\alpha+\beta)-1}
           \hat{q}^{\,-2\alpha-1}\hat{q}_\nu
{\cal P}_{\mu\nu}\\
\Gamma^{\rm tr(2)}_\mu&=&(-\gamma_\lambda\gamma_\rho\gamma_\nu
\hat{q}_\rho\hat{q}^{\,-2\beta}\hat{p}_\lambda
\hat{p}^{\,-2(\alpha+\beta)-1}\hat{q}^{\,-2\alpha-1}\nonumber\\
&&+i(D-2\alpha-2\beta-1)\gamma_\nu\hat{q}^{\,-2\beta}\hat{p}^{\,-2(\alpha+\beta)-1}
           \hat{q}^{\,-2\alpha-1}){\cal P}_{\mu\nu}\,.
\eea
These equations give
\bea
\langle x|\Gamma^{\rm tr}_\mu|y\rangle&=&\frac{i\Gamma(D/2-\alpha-\beta+1/2)}
                                   {\pi^{D/2} 2^{2\alpha+2\beta-1}\Gamma(\alpha+\beta+1/2)}
                  \nonumber\\
&&\times\Bigg(\delta_{\mu\nu}-
                     \frac{\partial{_\mu^y}\partial{_\nu^y}}{(\partial^2)^y}\Bigg)
\frac{\frac{1}{2}x^2(\rlap/x-\rlap/y)\gamma_\nu\rlap/y+
\frac{\beta}{D/2-\alpha-\beta-1/2}(x-y)^2\rlap/xy_\nu}
         {x^{2\beta+2}|x-y|^{D-2\alpha-2\beta+1}y^{2\alpha+1}}\,.          \label{app6}
\eea
Now we can put Eqs.\ (\ref{app1}),
(\ref{app4}) and (\ref{app6}) together to obtain an expression for the
position space matrix element of $\Gamma_\mu$ of Eq.\ (\ref{Gmu}).
{\it A crucial step is to perform $\partial{_\nu^y}$ in Eq.\ (\ref{app6}),
so that $\partial{_\mu^y}/(\partial^2)^y$ in Eq.\ (\ref{app6}) and in Eq.\ (\ref{app1})
can be taken together.}
(Whereas in Refs.\ \cite{my1} and \cite{my2}, the transverse and longitudinal parts were kept separate.) 
 We then end up with 
\bea
\frac{\partial{_\mu^y}}{(\partial^2)^y}
\frac{(2\alpha-1)(D-2\alpha-1)x^2 +2\beta((D-2\beta-2)y^2-2(2\alpha-1)x\cdot y)}
         {|x-y|^{D-2\alpha-2\beta+1}y^{2\alpha+1}}\,.                \label{app7}
\eea
in the expression for $\langle x|\Gamma_\mu |y\rangle$.
Now we use the relation 
\bea
(\partial^2)^y\frac{1}{|x-y|^m y^n}=\frac{n(n-D+2)x^2+(m+n-D+2)((m+n)y^2-2nx\cdot y)}
                                        {|x-y|^{m+2} y^{n+2}}\,.      \label{relation}
\eea
The expression (\ref{app7}) is therefore equal to $\partial{_\mu^y}
(1/(|x-y|^{D-2\alpha-2\beta-1}y^{2\alpha-1}))$. Note that {\it the conformal invariant
propagator $g_{\mu\nu}(x)/r^{D-2\beta}$ ensures that $\Gamma^{\rm tr}_\mu$
and $\Gamma^{\rm long}_\mu$ are added in the precise proportion so that $1/(\partial^2)^y$
can be taken care of.} After this, it is straightforward to arrive at Eq.\ (\ref{star2}).
\section{Some important relations}\label{app:imp}
The star-triangle relation involving three scalar fields is given by \cite{deramo, sym} 
\bea
\int d^Dx_4~ (x_{14}^2)^{-\delta_1} (x_{24}^2)^{-\delta_2} (x_{34}^2)^{-\delta_3}
&=&\pi^{D/2}\frac{ \Gamma(D/2-\delta_1) \Gamma(D/2-\delta_2) \Gamma(D/2-\delta_3)}{\Gamma(\delta_1)\Gamma(\delta_2)\Gamma(\delta_3)}\nonumber\\
&&\times (x_{12}^2)^{-D/2+\delta_3}
 (x_{13}^2)^{-D/2+\delta_2} (x_{23}^2)^{-D/2+\delta_1}\,,          \label{st}
\eea
where
\bea
\delta_1+\delta_2+\delta_3=D\,.                                        \label{delta}
\eea
The star-triangle relation for the Yukawa theory, involving two spinors and one scalar field,
is given by \cite{sym, fradkin1, my1}
\bea
&&\int d^Dx_4~ \frac{\rlap/x_{14}}{(x_{14}^2)^{\delta_1+1/2}}\,
              \frac{\rlap/x_{42}}{(x_{24}^2)^{\delta_2+1/2}}
               \frac{1}{(x_{34}^2)^{\delta_3}}\nonumber\\
&=&\pi^{D/2}\frac{ \Gamma(D/2-\delta_1+1/2) \Gamma(D/2-\delta_2+1/2) \Gamma(D/2-\delta_3)}
{\Gamma(\delta_1+1/2)\Gamma(\delta_2+1/2)\Gamma(\delta_3)}\nonumber\\
&&\times \frac{\rlap/x_{13}}{(x_{13}^2)^{D/2-\delta_2+1/2}}\,
          \frac{\rlap/x_{32}}{(x_{23}^2)^{D/2-\delta_1+1/2}}
           \frac{1}{(x_{12}^2)^{D/2-\delta_3}}\,                                   \label{st'}
\eea
where Eq.\ (\ref{delta}) holds again.
An analogous relation involving two scalars and one vector field is \cite{fradkin1}
\bea
&&\int d^Dx_4~ (x_{14}^2)^{-\delta_1} (x_{24}^2)^{-\delta_2} (x_{34}^2)^{-\delta_3}
g_{\mu\nu}(x_{14})\lambda_\nu^{x_4}(x_2x_3)\nonumber\\
&=&\pi^{D/2}(D-\delta_1-1)\frac{\Gamma(D/2-\delta_1) \Gamma(D/2-\delta_2) \Gamma(D/2-\delta_3)}
{\Gamma(\delta_1+1)\Gamma(\delta_2+1)\Gamma(\delta_3+1)}\nonumber\\
&&\times (x_{12}^2)^{-D/2+\delta_3+1}
 (x_{13}^2)^{-D/2+\delta_2+1} (x_{23}^2)^{-D/2+\delta_1}\,         
\lambda_\mu^{x_1}(x_2x_3)                          \label{st'''}
\eea
where $\delta_1+\delta_2+\delta_3=D-1$.
Eq.\ (\ref{st'''}) can be obtained by using the identity \cite{fradkin1}
\bea
g_{\mu\nu}(x_{14})\lambda_\nu^{x_4}(x_2x_3)=\frac{x_{12}^2}{x_{24}^2}\lambda_\mu^{x_1}(x_2x_4)
                                           -\frac{x_{13}^2}{x_{34}^2}\lambda_\mu^{x_1}(x_3x_4)
\eea
and the relation \cite{fradkin1}
\bea
&&\int d^Dx_4~ (x_{14}^2)^{-\delta_1} (x_{24}^2)^{-\delta_2} (x_{34}^2)^{-\delta_3}
\lambda_\mu^{x_1}(x_2x_4)\nonumber\\
&=&\pi^{D/2}\frac{\Gamma(D/2-\delta_1) \Gamma(D/2-\delta_2+1) \Gamma(D/2-\delta_3)}
{\Gamma(\delta_1+1)\Gamma(\delta_2)\Gamma(\delta_3)}\nonumber\\
&&\times (x_{12}^2)^{-D/2+\delta_3}
 (x_{13}^2)^{-D/2+\delta_2} (x_{23}^2)^{-D/2+\delta_1}        
\lambda_\mu^{x_1}(x_2x_3)\,,                                    \label{last}
\eea
where Eq.\ (\ref{delta}) holds.
[Eq.\ (\ref{last}), in turn, follows from Eq.\ (\ref{st}).]


\begin{thebibliography}{10}
\bibitem{footnote1}
In the process of amputing the spinor leg in
$C_{1\mu}^{d,l,\Delta}$, the structure $g_{\mu\nu}(x)/(x^2)^{\delta_3}$ automatically comes out
(see Eq.\ (\ref{eq:28})). So we need to have this propagator in the star-triangle relation instead
of the covariant gauge propagator. It may be mentioned that the transverse part of the covariant
gauge propagator can be directly related to the form $g_{\mu\nu}(x)/(x^2)^{\delta_3}$ only by a careful
regularization: see Eqs.\ (2.31)-(2.33) of Ref.\ \cite{fradkin2}.
\end{thebibliography}
\end{document}